# Comparative optic and dosimetric characterization of the HYPERSCINT scintillation dosimetry research platform for multipoint applications

Emilie Jean[1,2,3], François Therriault-Proulx[4] and Luc Beaulieu[1,2]

[1] Département de physique, de génie physique et d'optique et Centre de recherche sur le cancer, Université Laval, Quebec, QC, Canada
[2] Département de radio-oncologie et Axe Oncologie du CRCHU de Québec, CHU de Québec - Université Laval, Quebec, QC, Canada
[3] Département de radio-oncologie du CIUSSS-MCQ, CHAUR de Trois-Rivières, Trois-Rivières, QC, Canada
[4] Medscint inc. Quebec, QC, Canada

E-mail: Luc.Beaulieu@phy.ulaval.ca

**Abstract**

This study introduces the HYPERSCINT research platform (HYPERSCINT-RP100, Medscint inc, Quebec, Canada), the first commercially available scintillation dosimetry platform capable of multi-point dosimetry through the hyperspectral approach. Optic and dosimetric performances of the system were investigated through comparison with another commercially available solution, the Ocean Optics QE65Pro spectrometer. The optical characterization was accomplished by measuring the linearity of the signal as a function of integration time, photon detection efficiency and spectral resolution for both systems under the same conditions. Dosimetric performances were then evaluated with a 3-point plastic scintillator detector (mPSD) in terms of signal to noise ratio (SNR) and signal to background ratio (SBR) associated with each scintillator. The latter were subsequently compared with those found in the literature for the Exradin W1, a single-point plastic scintillator detector. Finally, various beam measurements were realized with the HYPERSCINT platform to evaluate its ability to perform clinical photon beam dosimetry. Both systems were found to be comparable in terms of linearity of the signal as a function of the intensity. Although the QE65Pro possesses a higher spectral resolution, the detection efficiency of the HYPERSCINT is up to 1000 time greater. Dosimetric measurements shows that the latter also offers a better SNR and SBR, surpassing even the SNR of the Exradin W1 single-point PSD. While doses ranging from 1 cGy to 600 cGy were accurately measured within 2.1% of the predicted dose using the HYPERSCINT platform coupled to the mPSD, the Ocean optics spectrometer shows discrepancies up to 86% under 50cGy. Similarly, depth dose, full width at half maximum (FWHM) region of the beam profile and output factors were all accurately measured within 2.3% of the predicted dose using the HYPERSCINT platform and exhibit an average difference of 0.5%, 1.6% and 0.6%, respectively.

## 1. Introduction

Plastic scintillators possess a unique set of advantages when it comes to clinical photon beam dosimetry as they offer small sensitive volume, water-equivalent material in terms of electronic density, a real-time response and one of the best energy independences relative to other dosimeters used in radiotherapy (Beddar *et al* 1992a, 1992b, Archambault *et al* 2006). Furthermore, the relationship between the scintillation light output and the dose deposition is directly proportional (Beaulieu and Beddar 2016). However, performing scintillation dosimetry using a multipoint plastic scintillator detector (mPSD) requires a highly sensitive array of photodetectors to collect the light emitted at different wavelengths from the scintillating elements and carried through a transport fiber (Andreo *et al* 2017, Therriault-Proulx *et al* 2012). The optical system also needs to offer sufficient spectral resolution to deconvolve various overlapping spectra (Archambault *et al* 2005, 2012). The company Medscint recently developed a scintillation dosimetry research platform (HYPERSCINT-RP100, Quebec, Canada), geared at optimizing the collection of light coming from plastic optical fibers (with numerical aperture of 0.5) for the purpose of multi-point scintillation dosimetry based on the hyperspectral approach (Archambault *et al* 2012, Therriault-Proulx *et al* 2012).

This study aims to characterize the HYPERSCINT$^{TM}$ scintillation dosimetry research platform through comparison with a commercial spectrometer QE65Pro (Ocean Optics, Dunedin, USA). The latter is a typical representative of the cooled spectrometer available on the market. It possesses a CCD detector array with high quantum efficiency designed for low-light level applications in the UV and visible range and is therefore suitable for scintillation dosimetry conditions. The two systems were compared with respect to essential optic and dosimetric properties, with the purpose of testing their performances under clinical photon beams while coupled to a mPSD. Results from Boivin et al. (Boivin *et al* 2015) of a commercially available Exradin W1 PSD (Standard Imaging inc., Middleton, USA), were also included in the study only for the dosimetric properties comparison as it relies on a two-channel photodiode enclosed in a shielded case providing triax outputs for green and blue signals only.

## 2. Materials and Method

The optic and dosimetric performances of the different systems were investigated using the same acquisition routine. For all measurements, lights were extinguished and black blankets covered the set-up to avoid signal contamination by ambient light. Background exposures with corresponding integration time were taken following each measurement and subtracted from the acquired signal. Throughout the experiment, the QE65Pro spectrometer was equipped with a 50 μm slit and the grating H3 (600 lines mm$^{-1}$ blazed at 500 nm) while the HYPERSCINT platform was connected with a 0.5 mm SMA905 adapter acting as the slit. The two of them were wavelength calibrated using a mercury-argon calibration source (HG-1, Ocean Optics, Dunedin, USA) prior to comparison while no correction were made to the intensity values. All measurements were performed using a wavelength range set from 400 to 700 nm which represents an effective area of 400 x 58 pixels on the detector of the QE65Pro. As for the HYPERSCINT, the wavelength range used represents an effective area of 3070 x 100 pixels. Since the Ocean Optics spectrometer is designed to automatically apply an on-chip binning on the 58 pixels forming each column of the photodetector array, a summation was also applied for the HYPERSCINT. Both detectors possess a cooling system that keeps the temperature stable across all measurements. That temperature was set at 10° C for the HYPERSCINT while that of the QE65Pro was set at -20° C.

### 2.1 Optical characterization

Optical measurements were performed with the Ocean Optics spectrometer and the HYPERSCINT platform directly connected to a calibration lamp to avoid any losses that would result from a transport fiber. The detection efficiency of both systems was measured using an intensity calibration lamp (HL-2000, Ocean Optics, Dunedin, USA) with fixed integration time in order to quantify intrinsic losses due to the slit size, the transmission of the optical elements and quantum detection efficiency of the sensor. As both systems offer a radiometric resolution of 16 bit per sampled wavelength, grayscale values were used to compare their sensitivity in terms of measured count per photon entering the system for a 1 second exposure. The surface power density of the calibration lamp was known for the wavelength range covered and allowed to quantify the number of photons emitted per second. The dependence of the integrated signal over the visible spectrum range as a function of the integration time from 10 ms up to 10 s was also studied using the same setup. For each integration time, three acquisitions were made, and both mean and standard deviation were calculated. Spectral resolution in the visible range was then evaluated with a mercury-argon wavelength calibration source (HG-1, Ocean Optics, Dunedin, USA) by measuring the full width at half maximum (FWHM) of the two prominent mercury lines located in the blue and green regions at 435.833 and 546.074 nm, respectively.

### 2.2 Dosimetric characterization

#### 2.2.1 mPSD fabrication and irradiation set-up

A previously optimized 3-point mPSD (Rosales et al 2019, 2020) composed of BCF60, BCF12 and BCF10 (Saint-Gobain

Crystals, Hiram, USA) respectively of 7, 6 and 3.5 mm long separated from each other by 1 cm of clear optical fiber (ESKA GH-4001, Mitsubitshi Chemical Co., Tokyo, Japan) was built to evaluate dosimetric performances. The total length of the mPSD assembly that includes the 3 scintillators and the clear fiber between them was 36.5 mm long. The same type of clear fiber was used to conduct the light emitted by the mPSD to the photodetector. As illustrated in figure 1, a green transmission filter placed after the BCF60 prevented cross-excitation from the BCF10 and BCF12 scintillators. An optical adhesive ensured all components of the mPSD to be bonded to each other. The scintillators, clear fiber and transport fiber had a 1 mm diameter. However, the mPSD was enclosed in a 2.2 mm diameter black jacket to prevent ambient light from being collected.

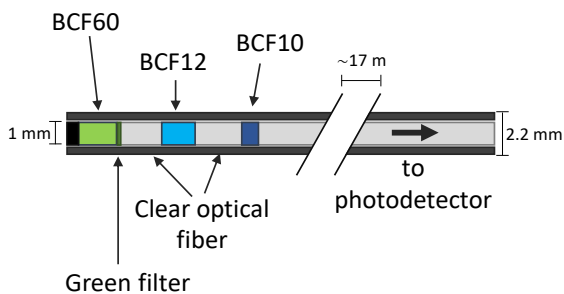

*Figure 1 : Design schematic of the mPSD showing the 3 scintillators used and the green filter position.*

All irradiations were performed with a Varian Clinac IX (Varian Medical Systems, Palo Alto, USA) linear accelerator (LINAC) under a 6 MV photon beam and a repetition rate of 600 MU/min. The probe was placed perpendicularly to the beam direction at a depth of 1.5 cm ($d_{max}$) in a solid water phantom and a 10 cm thick slab was used to provide backscatter. Both scintillation and Cerenkov signals were collected through a single 1 mm diameter transport fiber of 17 m long so the photodetectors could be placed outside the treatment room to minimize noise.

### 2.2.2 SNR and SBR

The dosimetric performances of the systems under a photon beam were quantify in terms of signal to noise ratio (SNR) and signal to background ratio (SBR) associated to each scintillator of the 3-point mPSD as a function of the dose rate. Those are defined as:

$$SNR = \frac{\mu_s}{\sigma_s}, \quad (1)$$

$$SBR = \frac{\mu_s}{\mu_b}, \quad (2)$$

where $\mu_s$ represents the mean signal for a determined irradiation in a fixed time, $\sigma_s$ the standard deviation of the same signal and $\mu_b$ the mean background signal.

While the mean signal depends on the sensitivity of the photodetector, the mean background will depend on the amplifier circuit configuration, performances of the cooling system, gain applied and permeability to light of the probe and the photodetector itself (Hopkinson *et al* 2004). According to the Rose criteria, proper detection of a signal strongly depends on the SNR and becomes possible when it exceeds five. Although signal can be detected while the SNR reaches lower values, performance degrades as it approaches zero (Bushberg 2002). As for the SBR, the threshold for a proper recognition of the signal was set at 2, meaning the signal must be at least twice the background intensity to provide sufficient contrast.

Dose measurements were achieved with each scintillator successively placed at the centre of a 4 x 4 cm$^2$ field size. The surface-to-source distance (SSD) was gradually increased in order to reduce the mean dose rate. For each dose rate, respectively five irradiations of 200 MU at 600 MU/min and background measurements were made. Each acquisition totalized 30 seconds using a repetitive 1 s integration time. Absolute dose rate measurements relative to SSD were performed with a TN 31014 ionization chamber (PTW, Freiburg, Germany) to account for the field size variation, attenuation in air, and treatment room backscattering at high SDD. The SNR and SBR were calculated for the two photodetectors using Eq. (1) and Eq. (2), respectively. The SNR of both systems were then compared to those of the W1 found in the literature (Boivin *et al* 2015). Since the W1 relies on a photodiode for the conversion of scintillation emission to electric signal (Hoehr *et al* 2018), the area under the curve of scintillation and background spectra in the visible range were used for the mean and standard deviation calculations.

### 2.2.3 Signal deconvolution and dose calibration

In order to determine the individual contribution of all scintillating elements and the Cerenkov light emission to the total signal collected, a deconvolution was performed using a hyperspectral approach (Archambault *et al* 2012). For a system of *n* scintillating elements, we assumed that the measured spectrum (*m*) is a linear superposition of the normalized spectra ($r_i$) of all the scintillators and Cerenkov light such as

$$m = \sum_{i=1}^{n+1} r_i x_i, \quad (3)$$

where $x_i$ represents the contribution of each light emission sources *i*. The *n+1* term accounts for Cerenkov signal that needs to be removed as it does not contribute to dose calculation.

In the present study, the sensor has *L* pixels considered as individual measurement channels to which are assigned a wavelength ($\lambda_j$). This can be expressed by a matrix such as

$$m = Rx$$

$$\begin{bmatrix} m_{\lambda 1} \\ m_{\lambda 2} \\ \vdots \\ m_{\lambda L} \end{bmatrix} = \begin{bmatrix} r_{1,\lambda 1} & r_{2,\lambda 1} & \cdots & r_{n+1,\lambda 1} \\ r_{1,\lambda 2} & r_{2,\lambda 2} & \cdots & r_{n+1,\lambda 2} \\ \vdots & \vdots & \vdots & \vdots \\ r_{1,\lambda L} & r_{2,\lambda L} & \cdots & r_{n+1,\lambda L} \end{bmatrix} \begin{bmatrix} x_1 \\ x_2 \\ \vdots \\ x_{n+1} \end{bmatrix}. \quad (4)$$

The left pseudo-inverse matrix method is used to solve this system for the variable $x$ as follows

$$x = (R^T R)^{-1} R^T m. \quad (5)$$

Solving Eq. (5) implies that one must first obtain the raw spectrum of each element of the probe such as they are affected by the response of both photodetectors. Scintillation spectrum of each component in absence of Cerenkov light were acquired with an XStrahl 200 orthovoltage unit (XStrahl LTD., Camberley, UK) at 120 kVp while Cerenkov light was acquired with a clear fiber under a 6 MV photon beam from a linear accelerator.

To calculate the dose received by each element of the probe, it is necessary to perform a calibration using an irradiation condition with a known dose for each scintillator ($d_{i,calib}$). Using Eq (5), the intensity calculated for this calibration irradiation ($x_{i,calib}$) will then be used to determine the dose received in any irradiation conditions ($d_i$) such as

$$d_i = d_{i,calib} \frac{x_i}{x_{i,calib}}. \quad (6)$$

### 2.2.4 Dose linearity

The relationship between the intensity collected and the dose deposited was measured by successively placing each scintillator at the centre of a 10 x 10 cm² field size. A dose calibration of the signal was accomplished beforehand under a reference condition, i.e., at a $d_{max}$ depth at the isocentre using 100 cGy repeated 5 times for each scintillator. The average obtained was therefore used to calculate the intensity collected per deposited dose. A dose varying from 1 cGy up to 600 cGy was then used to verify that the measured intensity matches the dose calibration.

### 2.3 PDDs, beam profiles and output factors

A second mPSD was built to realize PDDs, beam profiles and output factors measurements with the HYPERSCINT platform only. The probe was designed to offer a better spatial resolution than the initial one and was made using the same technique as described in the previous section. The scintillator sizes were respectively of 2, 1.5 and 0.8 mm long for the BCF60, BCF12 and BCF10 so the length ratios between them were kept identical as the first probe. The length of clear optical fiber used to separate the three scintillators was adjusted to obtain a center-to-center distance of 1 cm between each of them. The probe assembly length of 21.4 mm long was reduced compared to the first one. This ensured that more than one scintillator would be exposed in a 2 x 2 cm² field size for output factor measurements. A dose calibration of the signal was also performed with the new mPSD using the method described in the preceding section.

PDDs were acquired along the central axis of the beam at a SSD of 98.5 cm for a 10 x 10 cm² field size in a solid water phantom. Each scintillator was successively placed at the centre of the beam and PDDs were obtained by scanning each scintillator from a depth of 1.5 cm to 15 cm using various increments. Beam profiles were taken at a $d_{max}$ depth in a 10 x 10 cm² field size. The BCF12 scintillator was placed at the centre of the phantom and the couch was moved in the lateral direction by increment of 1 cm from -7 cm to 8 cm. For each couch position, doses were measured simultaneously by the three scintillators of the mPSD as a function of their respective lateral distances to the beam central axis. The set-up was then scanned using a computed tomography scanner (Somatom Definition, Siemens, Erlangen, Germany). The images were exported into the treatment planning system Pinnacle3 v9.8 (Phillips, Amsterdam, Netherlands) to predict the dose distribution in the phantom. Dose values measured with the mPSD for the PDDs and beam profiles were then compared with the predicted dose obtained from the TPS.

Similarly, output factors for field sizes varying from 2 x 2 to 15 x 15 cm² were measured with each scintillator of the mPSD successively placed at the isocentre using 3 repeated irradiations of 100 cGy. Relative output factors were defined as the ratio of the signal measured at field size $n \times n$ cm² using jaw-defines fields to the signal measured under a 10×10 cm² field. Those were then compared with a TN31014 pinpoint ionization chamber using the same set-up.

## 3. Results

### 3.1 Optic characterization

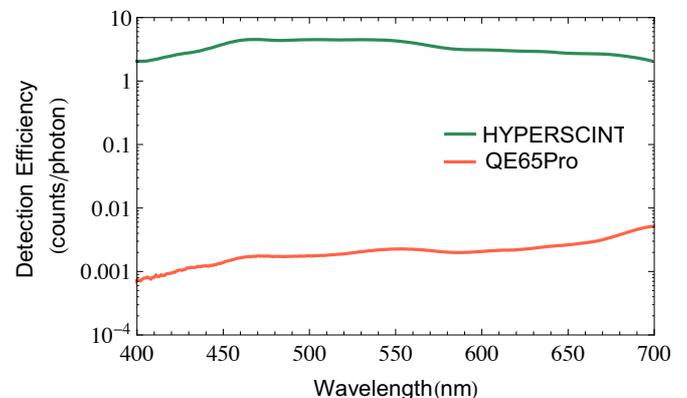

*Figure 2 : Detection efficiency in counts per photon entering the system as a function of the wavelength of the HYPERSCINT platform and the QE65Pro spectrometer.*

Comparison of the detection efficiency of both systems showed that the HYPERSCINT platform can detect up to 1000 times more photons than the Ocean Optics spectrometer for the same integration time as illustrated in figure 2. The sensitivity as a function of the wavelength was found to be stable for the HYPERSCINT while it was decreasing of a factor 10 in the blue region of the spectrum for the QE65Pro. This can be problematic as most of the scintillators commercially available tends to peak in the blue region.

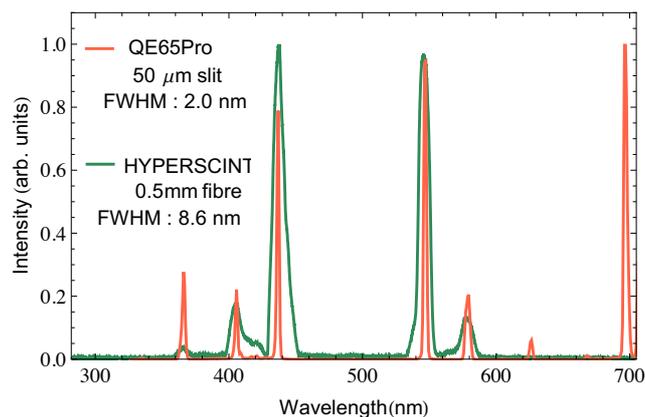

*Figure 3 : Spectra of the mercury-argon calibration source obtained with the HYPERCSINT platform and the QE65Pro normalized to their respective maximum. Corresponding FWHM were measured for the 2 prominent mercury lines.*

Using the same set-up, integrated signal over the visible spectrum wavelengths were obtained for integration times ranging from 10 ms up to 10 s. As expected, the collected intensity of both systems was found to follow a linear trend as a function of the integration time on the tested range. Still, spectral resolution of the QE65Pro was found to be more effective with a full-width-half-max of 2.0 nm for the 435.833 and 546.074 nm mercury lines while those of the HYPERSCINT reached 8.6 nm as shown in figure 3. As both spectra were normalized to their respective maximum, the response of the two systems as a function of the wavelength affects the relative intensity of the various lines.

### 3.2 SNR and SBR

Deconvolved spectra of the three scintillators of the mPSD and the Cerenkov contribution obtained with the calibration measurements can be seen in figure 4.a and figure 4.b for the QE65Pro and the HYPERSCINT, respectively.

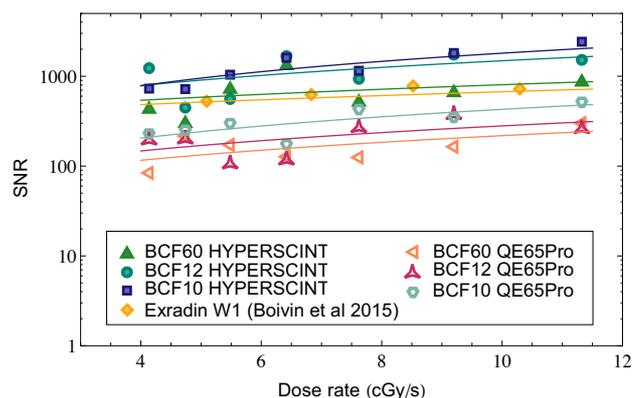

*Figure 5 : Signal to noise ratio associated to each scintillator of a 3-point mPSD as a function of the dose rate for the QE65Pro and the HYPERSCINT. The SNR of the single-point W1 PSD were obtained from the literature (Boivin et al 2015).*

The SNR obtained with the 3 scintillators of the mPSD as a function of dose rate are presented in figure 5. Each set of points is related by a power law fit to compare the various photodetectors used. The SNR of the QE65Pro was found to lie between 80 and 500 for dose rate in the range of 4 cGy/s to 11.5 cGy/s. Similarly, the HYPERSCINT SNR has reached values between 300 and 2300, where those of the BCF60 scintillator were found to be comparable to the Exradin W1 as illustrated in figure 5. Still, the latter were exceeded by the other two scintillators with up to a factor 4 greater signals.

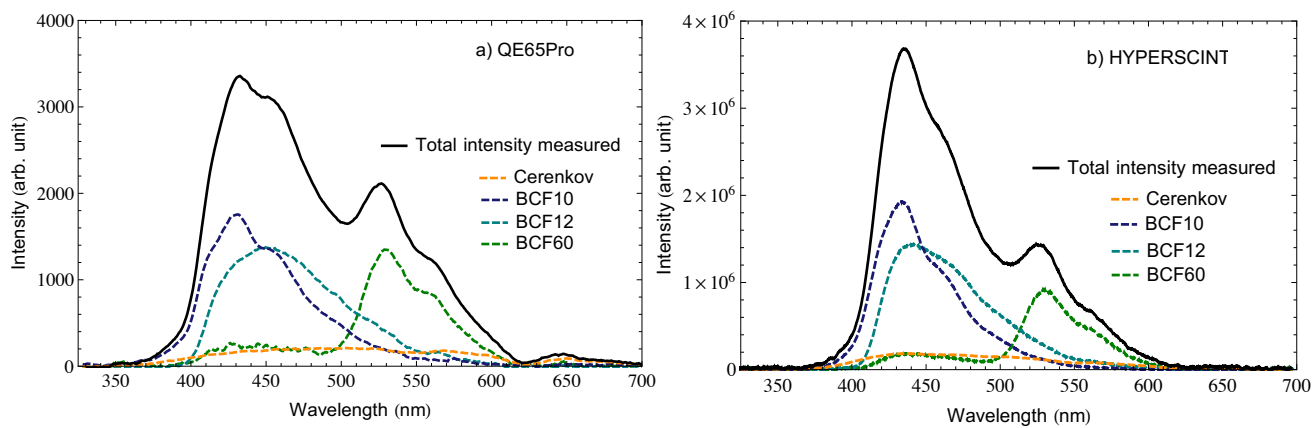

*Figure 4 : Total intensity measured for the calibration irradiation and deconvolved spectra of each light emitting element of the mPSD obtained with Eq. (5) for the QE65Pro (a) and the HYPERSCINT (b).*

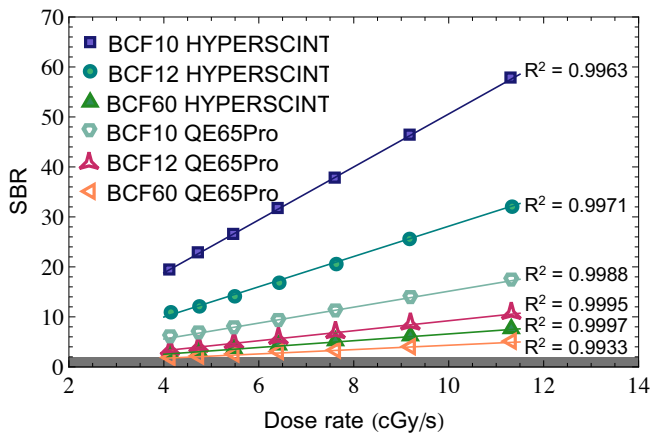

*Figure 6 : Signal to background ratio associated to each scintillator of a 3-point mPSD for 5 repeated 30 seconds acquisitions as a function of the mean dose rate for the QE65Pro and the HYPERSCINT. Grey area represent the detection threshold.*

Figure 6 illustrate the SBR of each scintillator of the mPSD as a function of the dose rate measured respectively with the QE65Pro and the HYPERSCINT. Those were calculated using the mean background values obtained from acquisition without irradiation and were respectively of 10.14 ± 2.94 and 8202 ± 412 count per second for the QE65Pro and the HYPERSCINT. The SBR of the former has reached a minimum of 1.6 for the BCF60 scintillator at 4 cGy/s while the latter did not fell under the detection threshold of 2 even at the lowest dose rate. As expected, the intensity of the scintillation signal collected as a function of the dose rate for fixed integration time was found to follow a linear trend for the two systems. While the linear dose-light relationship and the dose rate independence are well known properties of scintillators, this observation demonstrates that the linearity of the signal was preserved throughout the complete detection chain.

*3. 3 Dose linearity*

Signal characterization of the 3 scintillators using the HYPERSCINT platform displays a linear dose-light relationship for all the doses tested. In contrast, the linearity could not be achieved under 10 cGy with the QE65Pro. Figure 7.a shows that the latter was able to measure the dose accurately within ± 3.2% above 50 cGy while the maximum difference reached 86% for lower doses. The scintillation signal measured with the HYPERSCINT shows a discrepancy with the predicted dose that reaches 2.1% at 1 cGy. However, the difference falls within ± 0.7% above 10 cGy as illustrated in figure 7.b.

*3.4 PDDs, beam profiles and output factors*

The accuracy of the high spatial resolution mPSD coupled with the HYPERSCINT platform was validated by realizing various dose measurements. Depth dose along the central axis of the beam for a 10 x 10 cm$^2$ field size acquired with the mPSD in a solid water phantom are displayed in figure 8. The dose from the surface to 200 mm depth as predicted by the TPS is shown on the same graphic. A difference between the collected intensities and the predicted dose can be observed for the 3 scintillators of the mPSD. Nevertheless, the maximum deviation observed was 1.1%.

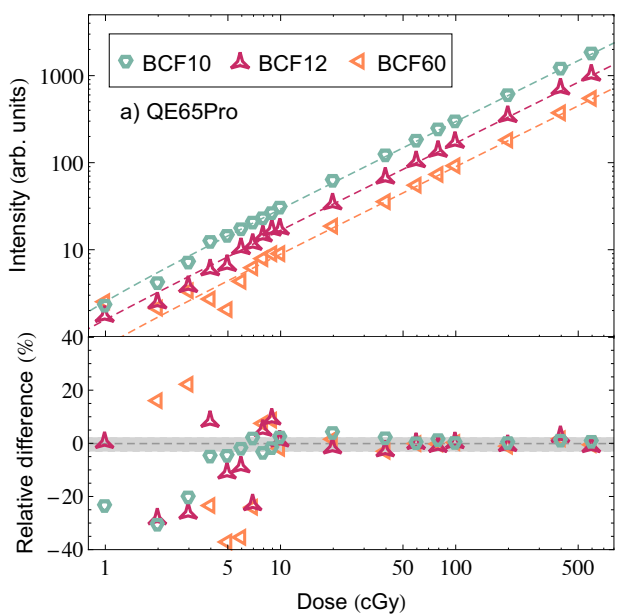
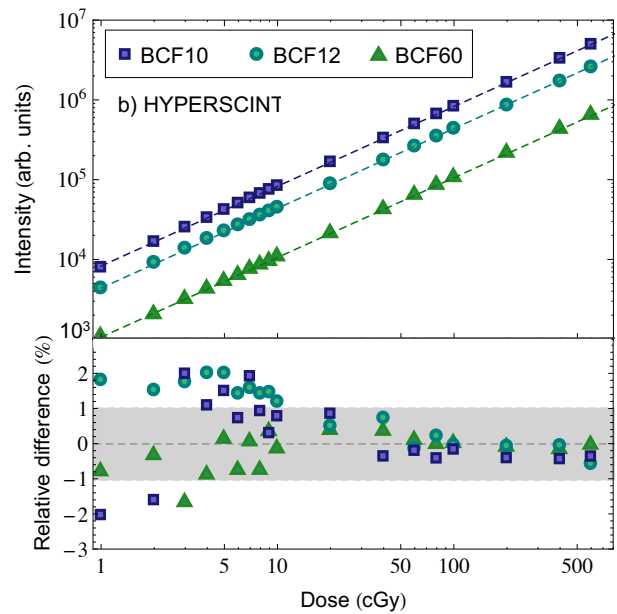

*Figure 7 : Signal measured from each scintillator of the mPSD as a function of dose for the QE65Pro (a) and the HYPERSCINT (b) and the relative difference between measured and predicted dose. The dashed lines represent the expected intensity as calculated with the calibration measurements.*

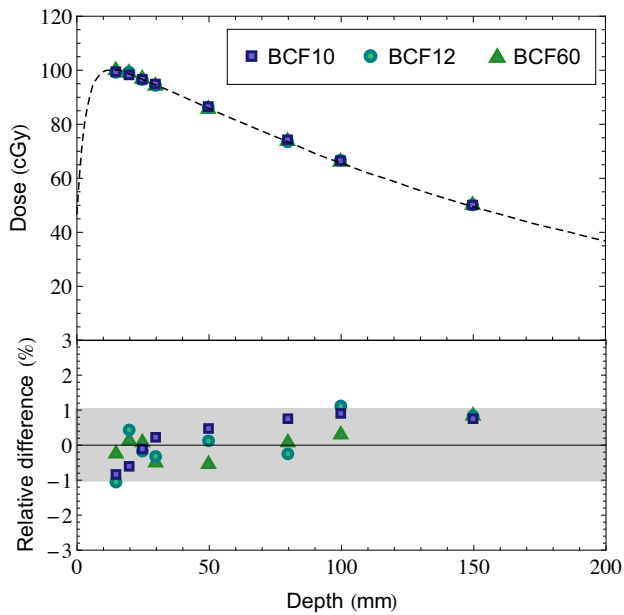

*Figure 8 : 10 x 10 cm² 6 MV depth dose measured with the mPSD compared to the TPS predicted dose (black dashed line) and their relative differences.*

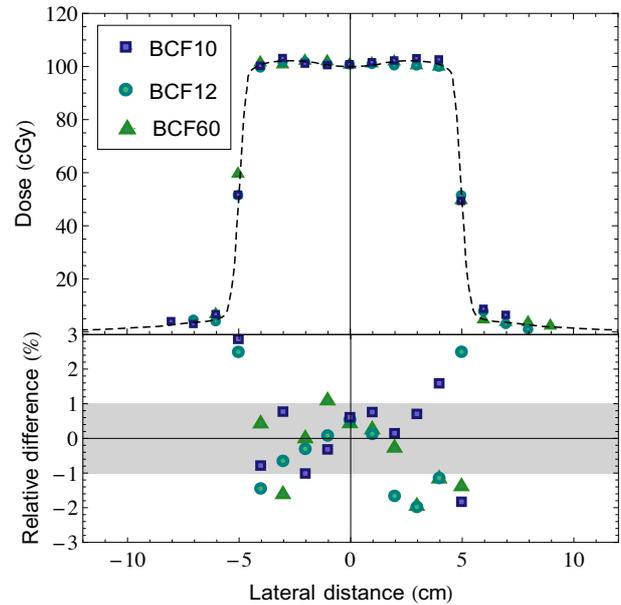

*Figure 9 : 10 x 10 cm² 6 MV beam profile measured with the mPSD compared to the TPS predicted dose (black dashed line) and their relative differences.*

Similarly, the full-width-half-max region of the beam profile was accurately measured with a maximum difference of 2.3% of the predicted dose, as illustrates figure 9. Somehow, the collected scintillation misrepresents the deposited dose in the penumbra region and the difference reaches up to 32% for the farthest points from the central axis.

Table 1 depicts relative output factors obtained with the mPSD for various field sizes normalized to the signal measured under a 10 × 10 cm² field. Comparison with an ionization chamber shows that measurements with the BCF12 and BCF60 scintillators exhibit an average difference of 0.7% and 1.1%, respectively. Best results were obtained with the BCF10 scintillator with an average difference of 0.4% and a maximal deviation of 0.5% for field size as small as 2 x 2 cm².

## 4. Discussion

### 4.1 Optical characterization

It is of interest to evaluate the optical performance of the photodetectors to verify that they are suitable for photon beam dosimetry in combination with a mPSD. Scintillation dosimetry with a multipoint probe requires a highly sensitive array of photodetectors in order to collect light emitted at different wavelengths by the scintillators.

*Table 1 : Relative output factors measured with the 3 scintillators of the mPSD in comparison with a TN31014 pinpoint ionization chamber and their relative differences.*

| Field size (cm²) | Ionization chamber | BCF60 | Relative difference (%) | BCF12 | Relative difference (%) | BCF10 | Relative difference (%) |
|---|---|---|---|---|---|---|---|
| 2 x 2 | 0.8762 | 0.8919 | -1.79 | 0.8839 | -0.87 | 0.8721 | 0.47 |
| 3 x 3 | 0.8997 | 0.9146 | -1.64 | 0.9051 | -0.60 | 0.9038 | -0.45 |
| 4 x 4 | 0.9233 | 0.9359 | -1.36 | 0.9308 | -0.81 | 0.9193 | 0.43 |
| 5 x 5 | 0.9419 | 0.9414 | 0.05 | 0.9482 | -0.67 | 0.9411 | 0.08 |
| 10 x 10 | 1 | 1 | - | 1 | - | 1 | - |
| 15 x 15 | 1.0324 | 1.0379 | -0.54 | 1.0368 | -0.42 | 1.0354 | -0.29 |

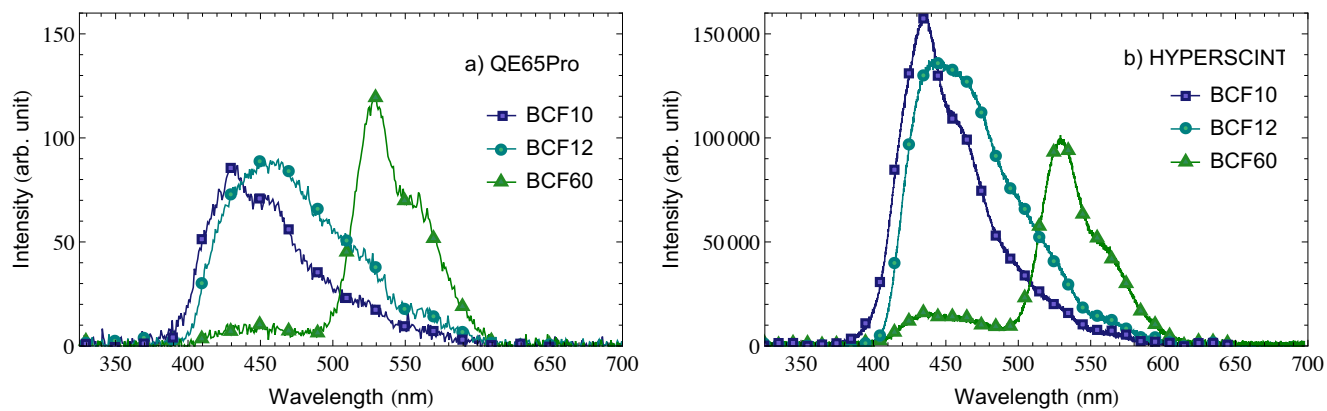

*Figure 10 : Scintillation spectra of each component of the mPSD in absence of Cerenkov light acquired with an XStrahl 200 orthovolage unit (XStrahl LTD., Camberley, UK) at 120 kVp with the QE65Pro (a) and the HYPERSCINT (b). The raw spectra are affected by the respective wavelength response of the two photodetection systems.*

Also, an adequate spectral resolution of the system is required to deconvolve multiple overlapping spectra. Both systems used in this study were comparable in terms of linearity of the signal as a function of the intensity. This property ensures that any signal variation is due to a change of the dose deposited in the scintillating elements. Moreover, the HYPERSCINT has proven to be far more sensitive than the QE65Pro which is a major asset for low dose measurements. Still, the latter has revealed a better spectral resolution than the HYPERSCINT platform. Sensitivity and spectral resolution of both systems are closely related to the slit sizes that were used. Nonetheless, the detection efficiency ratio of the two systems was found to be greater than the aperture ratio making the HYPERSCINT more sensitive for an equivalent amount of light entering the system. This can be linked to an optimized optical chain and the performances of the photodetector array itself. While the increased sensitivity is desirable for high spatial resolution mPSD and low dose rate measurements, it could induce signal saturation of some pixels at higher dose rates. However, the exposure time of the HYPERSCINT can be set as low as 30µs to overcome this possible issue.

*4.2 Dosimetric characterization*

The dosimetric characterization of the two photodetectors provides interesting elements to understanding the duality between light collection efficiency and spectral resolution. As demonstrated, sensitivity is the most significant property for dose measurement when using 3 scintillating elements with sufficient distinct spectra. In fact, scintillation spectra used to perform the deconvolution obtained with the orthovoltage unit were noisier with the QE65Pro as scintillators produced a relatively weak light emission at low dose rate as illustrated in figure 10.a. Increasing the integration time to gain signal was leading to a greater increase of the background level than the signal itself. Thus, the SBR for the 3 scintillators was approaching the detection limit of 2. This has led to considerable discrepancies in dose linearity measurements as the deconvolution accuracy strongly depends on the quality of the raw spectra. Furthermore, the poor results obtained for low doses and low dose rates can be linked to the total signal collected that was also weak. As for the HYPERSCINT, even if the background level is higher with $8202 \pm 412$ count per second in comparison with the $10.14 \pm 2.94$ count per second of the QE65Pro, a smaller standard deviation to the mean background ratio and a better light collection efficiency allows the acquisition of raw spectra of an appreciable quality (see figure 10.b). Moreover, the total signal collected for all doses and dose rates tested was sufficient to perform a proper deconvolution. Consequently, dose linearity measurements display only a slight difference above 10 cGy.

*4.3 Beam characteristics*

The beam characteristic measurements were only achieved with the HYPERSCINT platform due to low SNR that causes the impossibility to obtain raw spectra with the QE65Pro at the orthovoltage with the small field adapted mPSD. As the scintillating elements were smaller than the preceding version, the total intensity was weaker than the background level and statistical noise. Thus, the performances of the HYPERSCINT were evaluated in terms of clinically acceptable accuracy, which is $\pm$ 1% of the predicted dose.

Best results for the depth dose and output factor measurements were obtained with the BCF10 scintillator. This can be linked to the fact that this scintillator is the nearest to the photodetector and its emission is directly collected. On the other hand, the light emitted by the BCF12 and BCF60 must pass through multiple interfaces. For the FWHM region of the beam profile, the 3 scintillators of the mPSD have shown similar discrepancies while being greater than those obtained for the PDD and the output factors. Also, a greater difference was observed in the penumbra region, reaching up to 32%. According to the IAEA acceptance criteria (International Atomic Energy Agency 2007), the TPS prediction difference should not exceed 3 % of the deposited dose outside the beam edges for simple geometries such as opened square fields. Further investigations have demonstrated a variation of the linearity of the signal measured per wavelength as a function of the intensity. In fact, for a linear increase of the light intensity, the signal measured in the red region of the spectrum tends to increase faster than in the blue region. This is due to the summation process performed on the multiple pixels forming the columns of the photodetector array, each of them corresponding to a specific wavelength. As focus is achieved in the blue region of the spectrum, optical aberrations of the system degrade the performance in the green and red regions by spreading the light collected in a non-Gaussian shape along the Y axis of the detector. For low intensity measurements, the individual SNR of some pixels positioned away from the central axis falls under the detection limit. As a result, the sum underestimates the intensity for those wavelengths in such condition. Since not all optical aberrations are linear as a function of the intensity, the number of pixels reaching the detection limit shows a non-linear dependence. The fact that the probe emission spectrum shows a non-constant shape (red values are lower than blue) also contributes to this dependence when intensity decreases. The measured spectral shapes are then dependent of the total intensity collected while performing the calibration. This has led the algorithm used for the spectral deconvolution to underestimate the BCF60 intensity when the Cerenkov signal is greater and adversely underestimate the BCF10 when Cerenkov signal is weaker. While integrating over the whole spectrum for the optical characterization has lessen this effect, the variation of the Cerenkov light emitted along the beam profile emphasizes the problem. Reducing the effective range of pixel per column could provide better results but would also reduce the sensitivity. Replacing the post-processing summation by an on-chip binning of the pixels in the Y direction of the detector as done by the QE65Pro could solve this issue without compromising the sensitivity. However, this option was not available on the version of the system tested.

## 5. Conclusion

The purpose of this study was to characterize the HYPERSCINT scintillation dosimetry research platform through comparison with a commercial spectrometer QE65Pro. An optical characterization of the two systems has shown that the latter offers a better spectral resolution with a FWHM of only 2 nm while the former reached 8.6 nm. However, the HYPERSCINT platform offers a better light collection efficiency and allows to detect up to 1000 times more photons in its actual configuration. Furthermore, the dosimetric performances of the HYPERSCINT platform in terms of SNR individually measured for each scintillator of a 3-point mPSD has also surpassed the commercially available W1 PSD with up to a factor 4 greater signals for the BCF12 and BCF10. It also offers a SBR greater than the detection limit of 2 for all dose rate tested. As expected, the signal as a function of dose and dose rate was linear for the 3 scintillators. Besides, the accuracy of a high spatial resolution mPSD coupled with the HYPERSCINT platform was validated by measuring various beam characteristics. Depth dose, Full-Width-Half-Max region of the beam profile and output factors were all accurately measured within 2.3%.

Although the QE65Pro offers a better spectral resolution, the HYPERSCINT suits more effectively the needs for multipoint scintillation dosimetry by its sensitivity, SBR and SNR. This study constitutes a strong foundation for future applications of real-time scintillation dosimetry to external photon beam in radiotherapy.

## Acknowledgements

This work was supported by the *Natural Sciences and Engineering Research Council of Canada* (NSERC) Discovery grants RGPIN #2019-05038 and the Fellowships Program of the *Ministère de la Santé et des Services Sociaux du Québec* (MSSS). François Therriault-Proulx is Co-founder and CEO at Medscint inc., a company developing scintillation dosimetry systems. This work was not financially supported by Medscint.